\documentclass[3p,times,procedia]{elsarticle}
\usepackage{nupha_ecrc}

\volume{00}

\firstpage{1}

\journalname{Nuclear Physics A}

\runauth{M. Singh et al.}

\jid{nupha}

\jnltitlelogo{Nuclear Physics A}

\usepackage{amssymb}
\usepackage{amsthm}

\usepackage[figuresright]{rotating}

\begin{document}

\begin{frontmatter}



\dochead{XXVIIth International Conference on Ultrarelativistic Nucleus-Nucleus Collisions\\ (Quark Matter 2018)}

\title{Hydrodynamic Fluctuations in Relativistic Heavy-Ion Collisions}


\author[label1]{Mayank Singh}
\author[label2]{Chun Shen}
\author[label1]{Scott McDonald}
\author[label1]{Sangyong Jeon}
\author[label1]{Charles Gale}

\address[label1]{Department of Physics, McGill University, 3600 University Street, Montreal, QC, H3A 2T8, Canada}
\address[label2]{Physics Department, Brookhaven National Laboratory, Upton, NY 11973, USA}

\begin{abstract}
We present a novel approach to the treatment of thermal fluctuations in the (3+1)-D viscous hydrodynamic simulation MUSIC. The phenomenological impact of thermal fluctuations on hadronic observables is investigated using the IP-Glasma + hydrodynamics + hadronic cascade hybrid approach. The anisotropic flow observed in heavy-ion collision experiments is mostly attributed to the hydrodynamic response to the event-by-event collision geometry and to the sub-nucleon quantum fluctuations. However, hydrodynamic fluctuations are present during the dynamical evolution of the Quark Gluon Plasma (QGP) and are quantified by the fluctuation-dissipation theorem. They can leave their imprint on final-state observables. By analyzing the thermal noise mode-by-mode, we provide a consistent scheme of treating these fluctuations as the source terms for hydrodynamic fields. These source terms are then evolved together with hydrodynamic equations of motion. Such a treatment captures the non-perturbative nature of the evolution for these thermal fluctuations.

\end{abstract}


\end{frontmatter}


\section{Introduction}
\label{}

Relativistic viscous hydrodynamics has been very successful in explaining a large number of experimental observables in the heavy-ion collision program \cite{Ryu:2015vwa,McDonald:2016vlt}. The standard modelling of heavy-ion collisions uses an initial-state model followed by hydrodynamic expansion and hadronic cascade. Usually the only source of fluctuations included in the modelling is at the initial stage. Recently, interest in the effect of thermal hydrodynamic fluctuations has grown. The fluctuation-dissipation theorem states that fluctuations and dissipation go hand-in-hand. In this work, we implement the evolution of thermal fluctuations in MUSIC. Its effect on final state observables is studied.

Previous works have estimated the effect of thermal fluctuations in the context of simplified Bjorken\cite{Kapusta:2011gt} or Gubser flows\cite{Yan:2015lfa}. More realistic simulations have treated fluctuations as a perturbation\cite{Young:2014pka} or have used smearing of noise terms\cite{Sakai:2017rfi}. Building on those works, a systematic mode-by-mode analysis is done here to treat noise evolution within the hydrodynamic framework consistently. This technique could potentially be used for studying the critical fluctuations which are crucial in search for the critical point of the QCD phase diagram.

\begin{figure}
    \centering
\begin{tabular}{@{\hspace{0.0em}}c@{\hspace{0.1em}}c@{\hspace{0.0em}}}
    \includegraphics[width=0.45\textwidth]{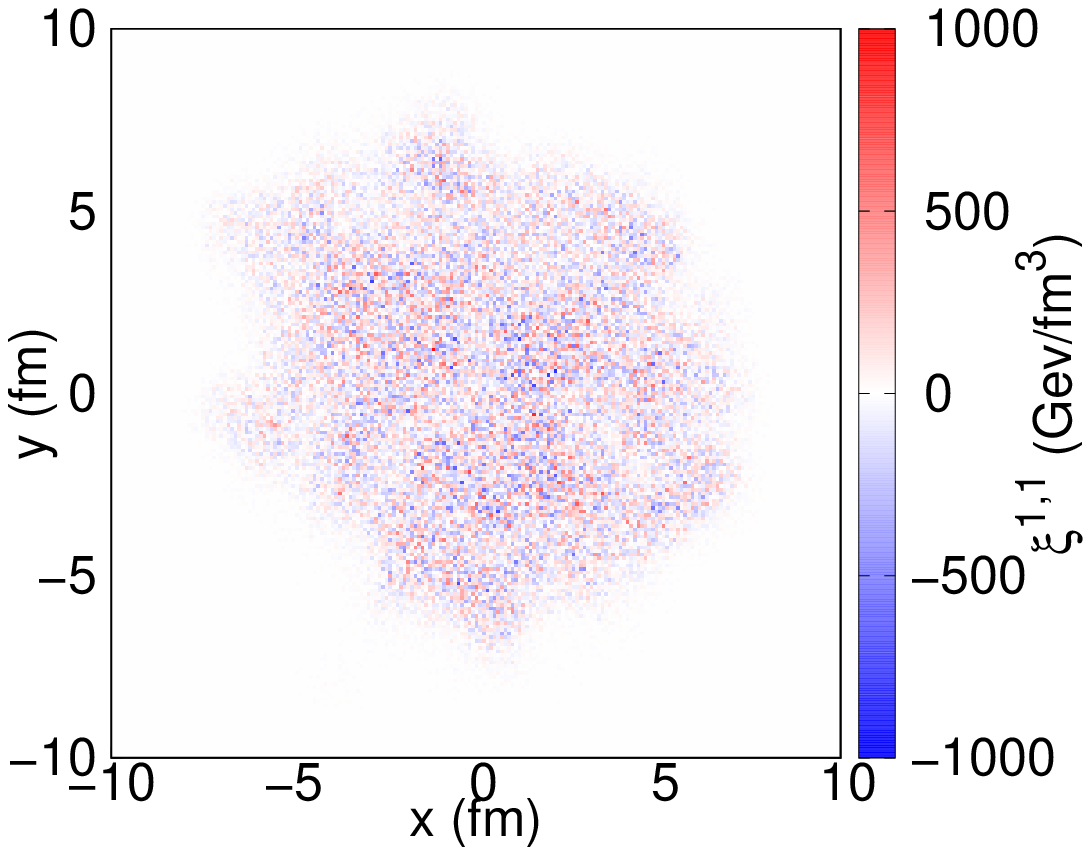} &
    \includegraphics[width=0.45\textwidth]{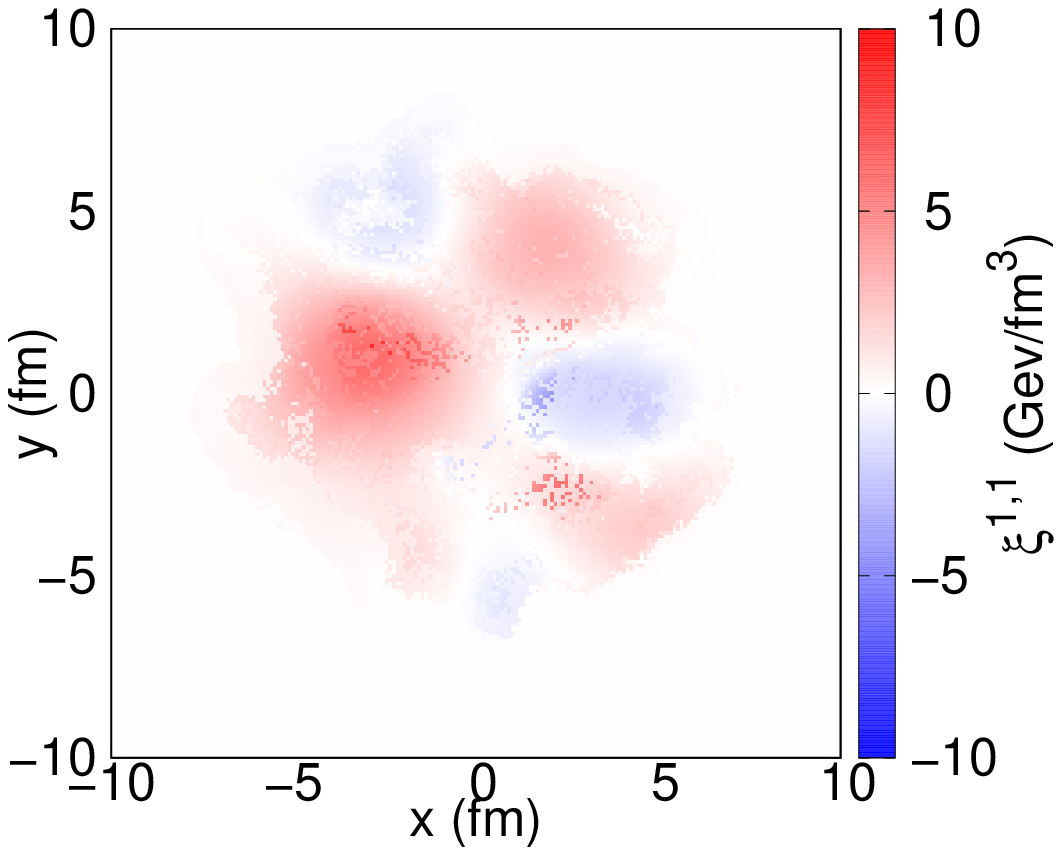}
\end{tabular}
    \caption{Sampled noise component $\xi^{1,1}$ at midrapidity. Left: before removing high wavenumber modes. Right: after removing modes larger than $ p_{cut} = 0.6/\tau_{\pi}$.}
    \label{fig1}
\end{figure}

\section{Implementation of stochastic source terms in the hydrodynamic framework}
Thermal fluctuations are introduced as a stochastic component of the energy-momentum tensor in fluid dynamics. The total energy-momentum tensor is conserved
\begin{equation}
    \partial_{\mu}(T^{\mu\nu}_{ideal} + \pi^{\mu\nu} + \Pi\Delta^{\mu\nu} + S^{\mu\nu}) = 0.
\end{equation}
Here, $T^{\mu\nu}_{ideal}$ is the ideal component of the energy-momentum tensor, $\pi^{\mu\nu}$ is the shear viscous term, $\Pi$ is bulk pressure and $\Delta^{\mu\nu} = g^{\mu\nu}-u^{\mu}u^{\nu}$ where $g^{\mu\nu}$ is mostly negative metric tensor and $u^{\mu}$ is the four-velocity. Both shear and bulk tensors have their evolution equations which, along with the equation of state, complete the system of equations for non-stochastic hydrodynamics.

The noise term $S^{\mu\nu}$ also has an evolution equation with a random stochastic term\cite{Young:2013fka, Chattopadhyay:2017rgh}
\begin{equation}
    (u\cdot\partial)S^{\mu\nu} = -\frac{1}{\tau_{\pi}}(S^{\mu\nu}-\xi^{\mu\nu} + \ldots).
\end{equation}
Here $\tau_{\pi}$ is the shear-relaxation time. Fluctuation term $S^{\mu\nu}$ can also be broken into two separate contributions with fluctuations associated with shear and bulk dissipation. In this work, only the contribution from shear-fluctuations is included. Bulk-fluctuations will be included in future work. Finally, the random noise term has an auto-correlation function given from the fluctuation-dissipation theorem\cite{Young:2014pka}
\begin{equation}\label{FDR}
    \langle \xi^{\mu\nu}(x)\xi^{\alpha\beta}(x')\rangle = 2\eta T\left[\Delta^{\mu\alpha}\Delta^{\nu\beta} + \Delta^{\mu\beta}\Delta^{\nu\alpha} - \frac{2}{3}\Delta^{\mu\nu}\Delta^{\alpha\beta}\right]\delta^{4}(x-x').
\end{equation}

Solving stochastic hydrodynamics could be tricky as one runs into issues with the noise term. The noise term is white i.e., noise in one cell is uncorrelated to noise elsewhere. This could result in arbitrarily large gradients in the system which is problematic for any PDE solver. Also noise contribution could be larger than the background itself. This could lead to negative energy densities which is clearly unphysical.

These issues can be avoided through the mode-analysis of the noise. The high frequency modes (wavenumber $>> \sqrt{\omega/\tau_{\pi}}$) decay on a very small time scale\cite{Akamatsu:2016llw} and are not relevant for this study. Here $\omega$ is the macroscopically relevant frequency.

The random noise term $\xi^{\mu\nu}$ is sampled from a Gaussian distribution, whose variance is specified in Eq. (\ref{FDR}). Use of Gaussian random numbers is justified because of the Central-Limit Theorem. We take the Fourier transform of the sampled noise using the FFTW package\cite{FFTW05}. Cutoff wavenumber $p_{cut}$ is determined locally in each cell. For this study, $p_{cut} = 0.6/\tau_{\pi}$. Noise is inverse Fourier transformed in each cell with all modes above $p_{cut}$ set to zero. Figure \ref{fig1} shows one sampled component before and after high mode removal.

The framework and transport coefficients used are same as our previous study without thermal fluctuations \cite{McDonald:2016vlt}.


\section{Results}
\begin{figure}
    \centering
    \includegraphics[width=0.450\textwidth]{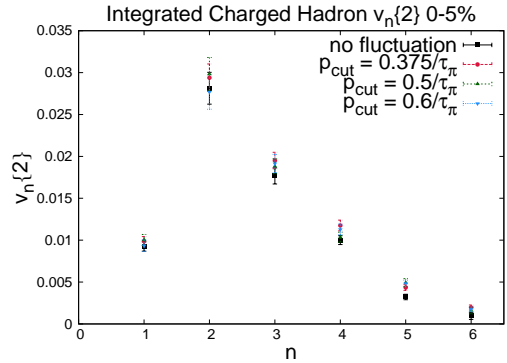}
    \caption{Charged hadron $v_{n}$ for different $p_{cut}$ for Pb-Pb collision at $\sqrt{s} = 2.76$ TeV.}
    \label{fig2}
\end{figure}

First, it needs to be tested if the results are indeed $p_{cut}$-independent. Different values of $p_{cut}$ are used and the results for $v_{n}$ are plotted in figure \ref{fig2} as a function of $n$. There is statistically significant effect on $v_{5}$ and $v_{6}$. Here the increment due to fluctuations is $p_{cut}$ independent which points to the fact that high wavenumber modes are indeed decaying fast. Having more effect on higher flow harmonics is also reasonable as those are more dependent on the small scale fluctuations, the kind being studied here.

\begin{figure}
    \centering
\begin{tabular}{@{\hspace{0.0em}}c@{\hspace{0.1em}}c@{\hspace{0.0em}}}
    \includegraphics[width=0.45\textwidth]{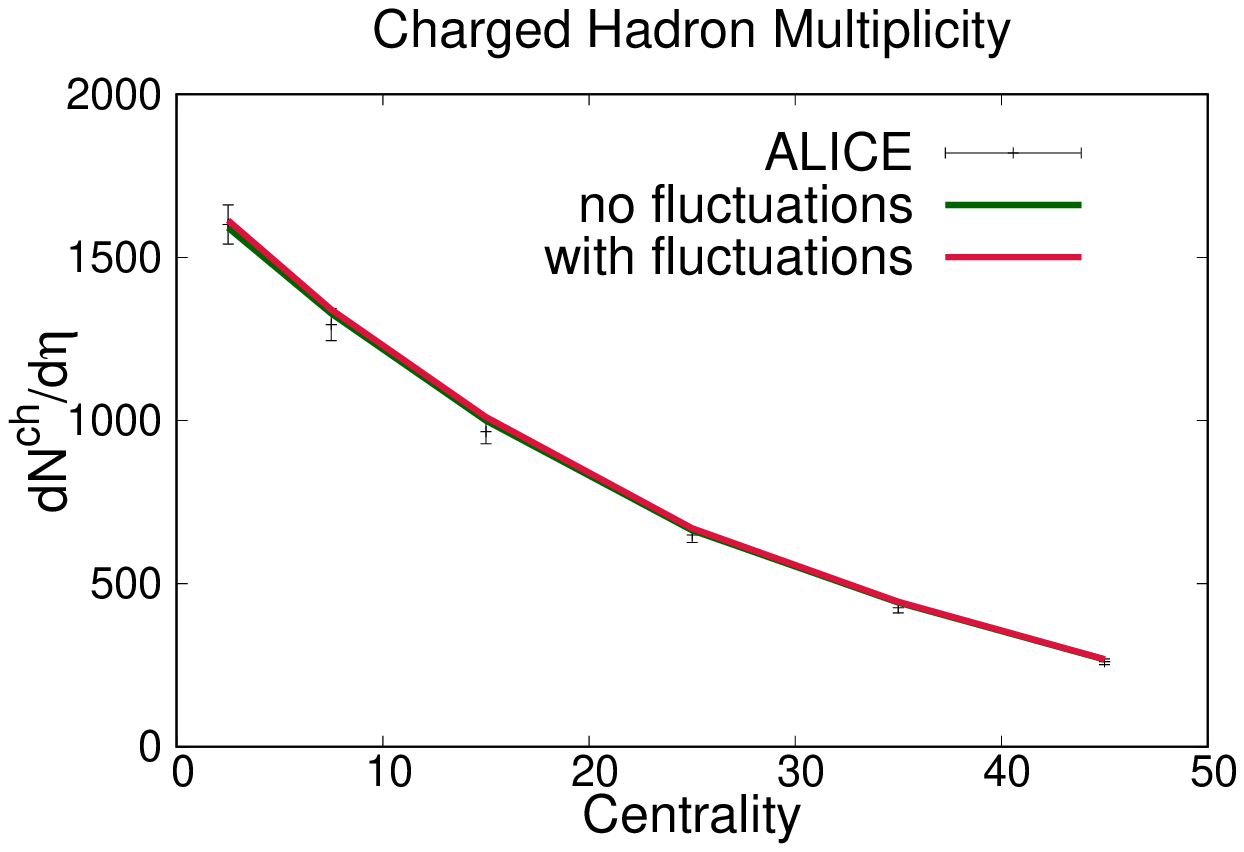} &
    \includegraphics[width=0.45\textwidth]{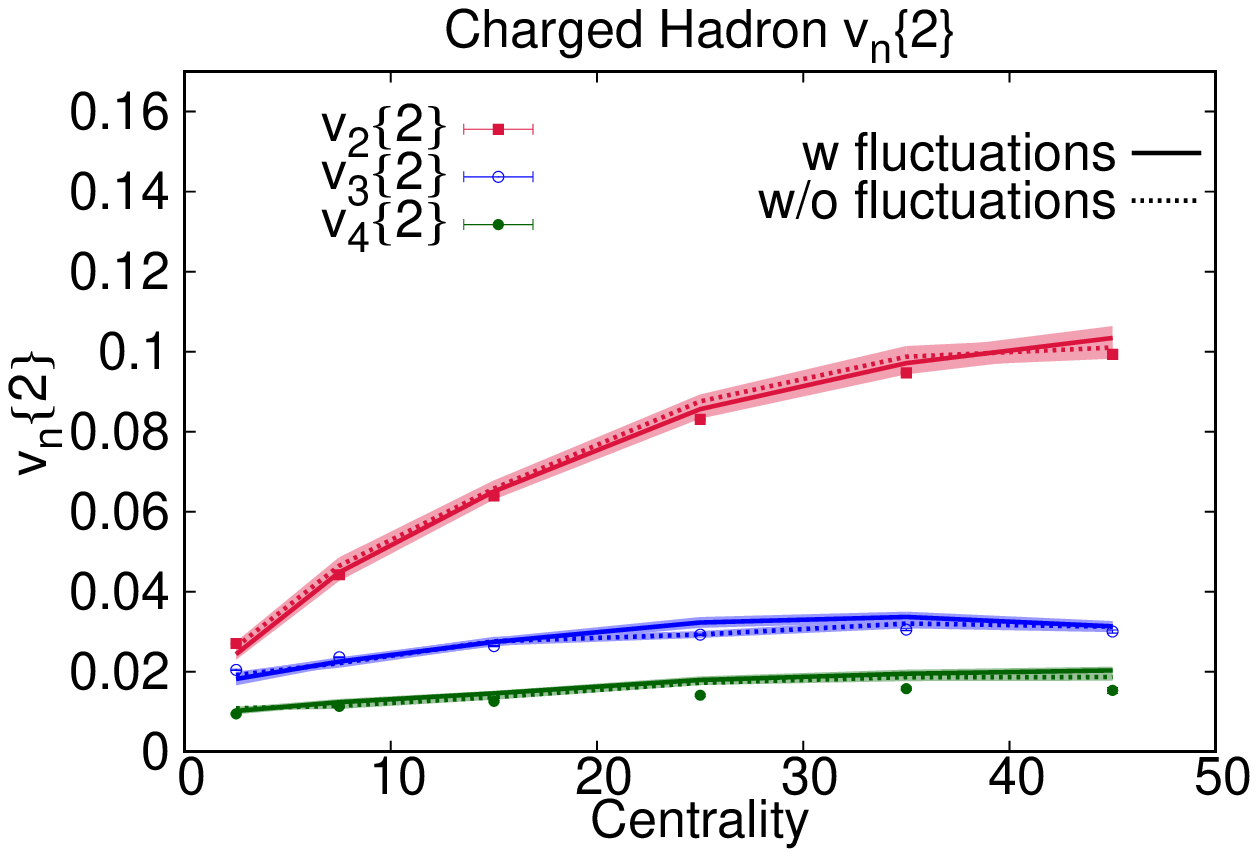}
\end{tabular}
    \caption{Left: Charged hadron multiplicity, Right: Charged hadron $v_{n}$, for Pb-Pb collision at $\sqrt{s} = 2.76$ TeV compared to data from ALICE collaboration \cite{Aamodt:2010cz,ALICE:2011ab}.}
    \label{fig3}
\end{figure}

Next the particle multiplicities and their $v_{n}$ as a function of centrality is shown in figure \ref{fig3}. As one can see, multiplicity is unaffected by the thermal fluctuations. Thermal fluctuations can cause entropy production (and by extension the multiplicity) to fluctuate around a mean value. However, their ensemble averaged value is unaffected\cite{Nagai:2016wyx}.

There is no statistically significant effect of thermal fluctuations on integrated charged particle $v_{2}$, $v_{3}$ and $v_{4}$. This implies that the fluctuations from the initial state IP-Glasma are more prominent at these centralities as far as these observables are concerned.

\begin{figure}
    \centering
    \includegraphics[width=0.450\textwidth]{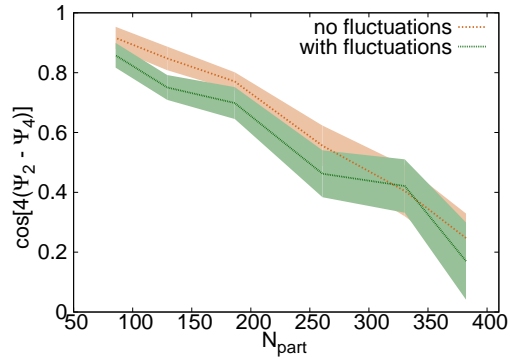} 
    \caption{Event plane correlator between $v_{2}$ and $v_{4}$ for 0.5 GeV $\leq p_{T} \leq$ 2.0 GeV for Pb-Pb collisions at $\sqrt{s} = 2.76$ TeV.}
    \label{fig4}
\end{figure}

Next event-plane correlators are considered which are more susceptible to small scale fluctuations. The event plane correlator $\cos[4(\Psi_{2}-\Psi_{4})]$ is plotted in figure \ref{fig4}. Even though values of $v_{2}$ and $v_{4}$ do not change much with thermal fluctuations, their respective planes are more decorrelated. This is in line with the previously known result that the thermal fluctuations decorrelate flow angles between different $p_{T}$ bins \cite{Sakai:2017rfi}.

\section{Discussions and Outlook}
A new implementation of thermal fluctuations in hydrodynamic simulations of QGP is presented. For Pb-Pb collisions at $\sqrt{s} = 2.76$TeV, thermal fluctuations do not affect lower flow harmonics but they affect their correlations. In future, more such small scale fluctuation sensitive observables such as 
the distributions of multiplicity and $v_{n}$ will be evaluated. Recently, experimental collaborations have measured higher flow harmonics which again carry a strong signature of thermal fluctuations. Stochastic hydrodynamic simulations would give better phenomenological understanding of these. Using non-boost invariant initial conditions, rapidity dependence of two-particle correlation could be calculated. Bulk fluctuations could be treated in a similar fashion and will be included in future.

This new approach could be used to revisit several of the physical parameters associated with the different epochs of the collision history,  particularly the QGP phase. The fluctuation strength is a function of temperature. So, the larger fluctuations at earlier times in the evolution will then leave their signature on those observables which decouple from the system then. 

With this technique in place, small systems could potentially be studied. There thermal fluctuations are expected to be more dominant\cite{Yan:2015lfa}. This could also possibly be used to look at large fluctuations near the critical point. 

\section{Acknowledgements}
We would like to thank Derek Teaney and Bjoern Schenke for useful discussions. This work is supported in part by the Natural Sciences and Engineering Research Council of Canada. Computation for this work was done in part on the supercomputer Cedar maintained by WestGrid and Compute Canada and on supercomputer Guillimin managed by Calcul Quebec and Compute Canada. MS and SM are supported in part by the Bourses d'excellence pour $\acute{\mathrm{e}}$tudiants $\acute{\mathrm{e}}$trangers (PBEEE) from Le Fonds de recherche du Qu$\acute{\mathrm{e}}$bec - Nature et technologies (FRQNT).  CG is grateful to the Canada Council for the Arts for funding through its Killam Research Fellowship Program.




\bibliographystyle{elsarticle-num}
\bibliography{references}







\end{document}